\documentclass[fleqn,10pt]{wlscirep}
\usepackage[utf8]{inputenc}
\usepackage[T1]{fontenc}
\usepackage{hyperref}
\usepackage{multirow}
\usepackage{graphicx}  % For including graphics
\usepackage{subcaption} % For subfigures
\usepackage{booktabs}
\usepackage{csvsimple} % To read CSV files
\usepackage{tikz}
\usetikzlibrary{positioning}
\usepackage{longtable}
\title{The Northeast Materials Database for Magnetic Materials}

\author[1]{Suman Itani}
\author[1,2]{Yibo Zhang}
\author[1,*]{Jiadong Zang}

\affil[1]{Deparment of Physics and Astronomy, University of New Hampshire, Durham, 03824, USA}
\affil[2]{Department of Chemistry, University of New Hampshire, Durham, 03824, USA}

\affil[*]{jiadong.zang@unh.edu}

\begin{abstract}

The discovery of magnetic materials with high operating temperature ranges and optimized performance is essential for advanced applications. 
Current data-driven approaches are limited by the lack of accurate, comprehensive, and feature-rich databases. This study aims to address this challenge by using Large Language Models (LLMs) to create a comprehensive, experiment-based, magnetic materials database named the Northeast Materials Database (NEMAD), which consists of 67,573 magnetic materials entries(\href{http://www.nemad.org}{www.nemad.org}). The database incorporates chemical composition, magnetic phase transition temperatures, structural details, and magnetic properties. Enabled by NEMAD, we trained machine learning models to classify materials and predict transition temperatures. Our classification model achieved an accuracy of 90\% in categorizing materials as ferromagnetic (FM), antiferromagnetic (AFM), and non-magnetic (NM). The regression models predict Curie (N\'eel) temperature with a coefficient of determination (R$^{2}$) of 0.87 (0.83) and a mean absolute error (MAE) of 56K (38K). 
 These models identified 25 (13) FM (AFM) candidates with a predicted Curie (N\'eel) temperature above 500K (100K) from the Materials Project. This work shows the feasibility of combining LLMs for automated data extraction and machine learning models to accelerate the discovery of magnetic materials.

\end{abstract}

\begin{document}

\flushbottom
\maketitle

\section*{Introduction}
 
For centuries, magnetic materials have been discovered and studied due to their broad applications in modern science and technology, including data storage devices, energy technologies, advanced medical equipment, quantum computing, and consumer electronics\cite{coey2010magnetism, gutfleisch2011magnetic, jungwirth2018multiple}. Searching for efficient magnetic materials is crucial to addressing global energy challenges and revolutionizing the technology industry\cite{skokov2018heavy}. For example, high-performance permanent magnets can increase efficiency in renewable energy, like wind power and hydroelectric power generators. At the same time, the consumption of fossil fuels and greenhouse gases is reduced. It can also give us high
density data storage solutions. Despite of these wide applications, limits for existing magnetic materials are present. For example, most high-performance magnetic materials contain rare earth elements and have a limited operating temperature range \cite{coey1997permanent, coey2020perspective}. The discovery of novel magnetic materials with greater operating temperature ranges using more abundant elements is a fundamental challenge in material science, partly due to the vast combinatorial space of possible compositions and the limitations of conventional methods.

The conventional techniques for material discovery have been largely based on systematic exploration of compositional space and intuition-directed experimentation. Although successful, these approaches require a lot of time and resources, and they can take years to produce useful outcomes \cite{himanen2019data}. The development of computational methods has opened up exciting new avenue toward materials discovery and property prediction. The first principles calculation such as the density functional theory (DFT) has enabled the predictions of material properties\cite{kohn1965self,jain2016computational}. Exploration of vast compositional spaces has been possible using high-throughput computational screening \cite{jain2013commentary,curtarolo2012aflow}. These computational breakthroughs have resulted in the construction of huge materials databases, such as the Materials Project \cite{jain2013commentary}, AFLOW \cite{curtarolo2012aflow}, AFLOWLIB \cite{curtarolo2012aflowlib}, and OQMD \cite{saal2013materials}.
Unfortunately, applying DFT to predict the magnetic properties of magnetic materials generally leads to less reliable results. Using the standard exchange-correlation functionals could not accurately describe the strongly correlated electron system in magnetic materials especially itinerant magnets. To accurately describe such a system, one should impose the space and spin symmetry constraints as done in ab initio wave function theory. However, the DFT  method could not properly incorporate these constraints\cite{illas2006spin}. The integration of electronic structure information with mean-field models is a standard approach for estimating magnetic properties. This methodology is commonly used to determine exchange interactions and construct an effective Hamiltonian when calculating the Curie temperature.\cite{mimura1978magnetic,yano2000molecular,gangulee1978mean,turek2005electronic}. This method, however, receives good results only for a short list of materials\cite{turek2006exchange}. Moreover, the requirement for prior knowledge of the crystal structure of the material also puts a limit on the applicability of the DFT method. 
Another drawback of DFT calculation is that it is mostly limited to the materials of small unit cells due to the high computational costs of large unit cells\cite{schuch2009computational,cohen2008insights}. Due to the above constraints, determining the magnetic properties and discovering novel magnetic materials remains a difficult task.

The data-driven approach is a new paradigm of research that has received great interest in recent years because of its capabilities in accurate predictions and discovering hidden patterns that otherwise may get missed by traditional methods\cite{butler2018machine,ramprasad2017machine}. Machine learning predicts physical properties quickly, discovers new materials, and optimizes the existing ones by analyzing large datasets\cite{badini2023unleashing,spear2018data, sanvito2017accelerated}. Specifically, several machine learning models have been developed to classify the magnetic material and predict magnetic properties, including phase transition temperatures and coercivity fields\cite{nelson2019predicting,court2020magnetic,singh2023physics,acosta2022machine,long2021accelerating,belot2023machine,lu2022fly,nguyen2019regression,choudhary2023machine,bhandari2024accurate,merker2022machine,xu2024predicting}. However, the performance of all these machine learning models highly depends on the availability and quality of a comprehensive magnetic materials database, which has not yet been fully available. It is rather challenging to identify trends across different classes of magnetic materials without such a comprehensive database, potentially overlooking promising research directions. For instance, the models were trained on extremely small data points of high-temperature regions specifically for the phase transition temperature prediction\cite{nelson2019predicting,belot2023machine}. This may generate less accurate predictions in that region which in turn undermines the discovery of high transition temperature materials. 
MAGDATA \cite{gallego2016magndata1,gallego2016magndata2} is a manually curated database comprising information for around 2000 magnetic materials with experimentally verified lattice and magnetic structure features. Due to the dataset's modest size, fewer features and single-entry lookups, the implementation of data-driven methods on that is challenging and less effective. Some articles and books have noticed some efforts to create a magnetic material database, but the data are scattered and hard to access electronically and systematically \cite{nelson2019predicting,singh2023physics,xu2011inorganic,connolly2012bibliography,buschow2003handbook}.

With the advent of the natural language processing (NLP), its combination with machine learning makes it possible to generate an automated database of magnetic materials.
An automatically generated database using the ChemDataExtractor toolkit and Snowball algorithm shows the power of this technique by including 39,822 records\cite{court2018auto}. But the database is restricted to only a few features, such as chemical compositions and their associated magnetic phase transition temperatures, which are relatively easy to extract from articles. This approach was less accurate when extracting information involving multiple materials or when the data came in varied sentence structures.
Another study has been done to build an auto-generated database of 3,613 ferromagnetic compounds with their corresponding Curie temperature\cite{gilligan2023rule}. They used a method based on fine-tuning bidirectional encoder representations from transformers (BERT) models. The database shows high quality and accuracy, making it valuable for developing machine learning models to predict material properties. However, database size and features are limited, and they failed to  extract data from the tables of a scientific article. Nevertheless, the machine learning models built on both auto-generated databases showed encouraging performance, as indicated by the statistical parameters R$^{2}$, MAE, and RMSE.

However, the magnetic properties of the materials strongly depend on the structural details of the material such as crystal structure, lattice structure, lattice parameters, and space group symmetry\cite{blundell2001magnetism}. Also, for the study of a particular magnetic material such as a permanent magnet, one needs to know the information about magnetization, coercivity, and magnetic anisotropy. All of the existing auto-generated databases lack these important information. Therefore, to make the data-driven method more effective for magnetic material discovery, a comprehensive database with greater features is needed.

Recent advancement of NLP, the large language models (LLMs), enables researchers to quickly and easily interact with a chatbot to extract their desired information in a structured format from a large chunk of unstructured text. Its applications in the field of science and technology are emergent\cite{dagdelen2024structured,polak2024extracting,yuan2024enhancing}. As one important application, the GPTArticleExtractor workflow \cite{zhang2024gptarticleextractor} developed by the authors provides a way to make a comprehensive material property database by automatically extracting data from scientific articles with high precision.

In this work, using the workflow shown in Figure 1, we have developed a database of 67,573 magnetic materials, incorporating chemical composition, structural details (such as
crystal structure, lattice structure, lattice parameters, and space group symmetry), and magnetic properties (such as coercivity,
magnetization, magnetic moment, remanence, and susceptibility). 
We then implemented this database to train machine learning models, including Random Forest (RF) classifier, XGB Classifier, Random Forest (RF) regressor, Extreme Gradient Boosting (XGBoost), and Ensemble Neural Network (ENN) for classifying magnetic materials and predicting transition temperature (Curie and N\'eel). The performance of these models were evaluated by cross-validation techniques. 
Our classification model achieved 90\% accuracy in classifying materials as ferromagnetic, antiferromagnetic, or non-magnetic. Our best Curie temperature prediction model with features generated from chemical composition (chemical composition + structure ) predicts Curie temperature with an R$^{2}$ value of 0.87 (0.83) and a mean absolute error of 56K (52K). Similarly, our best N\'eel temperature prediction model with features generated from only chemical composition predicts N\'eel temperature with an R$^{2}$ value of 0.83 and a mean absolute error of 38K. This model identified 25 promising ferromagnetic candidates with predicted Curie temperatures exceeding 500K and 13 antiferromagnetic compounds with predicted N\'eel temperature greater than 100K.

\begin{figure}[ht]
\centering
\includegraphics[width=0.9\linewidth]{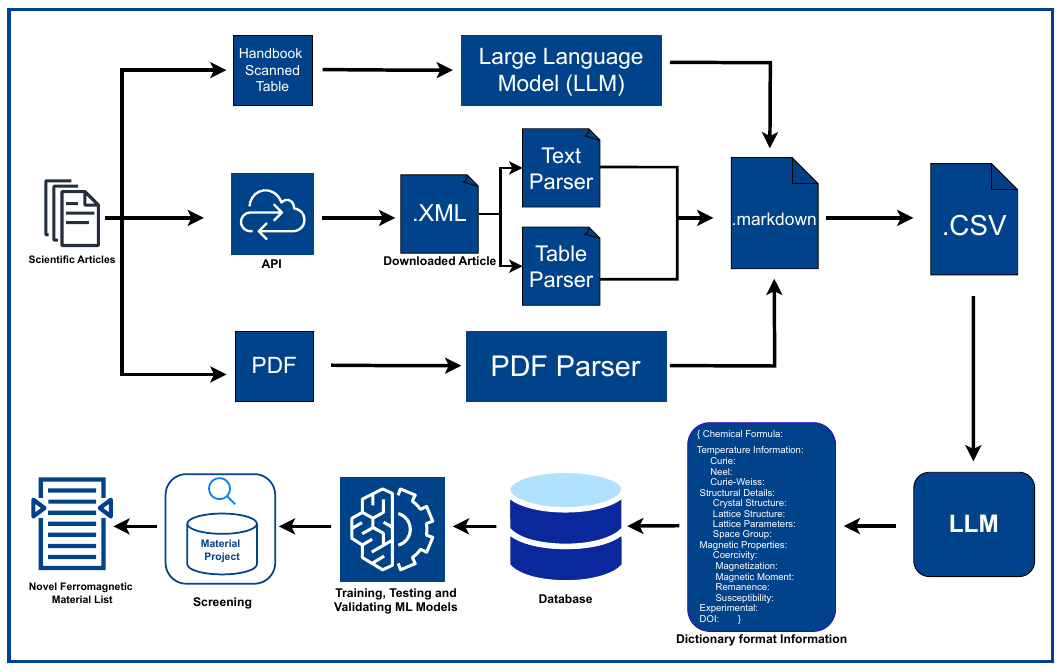}
\caption{\textbf{Workflow for the Construction and Analysis of a Magnetic Materials Database.} Scientific articles are processed via three pathways based on their format. Articles retrieved through the Journal's API in XML format are parsed using both a text parser and a table parser. Standard PDF documents are handled by a PDF parser, which converts the content into markdown text. For older, scanned or image based PDFs and historical handbooks, we use Google Gemini’s OCR capabilities to extract text and tables accurately. For longer documents like handbooks, the content is processed page-by-page and converted into markdown format. All markdown outputs are then converted into CSV files. These files are passed through GPT-4o with structured prompts to extract relevant materials data in a consistent JSON format. After cleaning and standardizing the extracted information, we compile it into the NEMAD database. The curated dataset is used to train machine learning models for classification and for predicting Curie and N\'eel temperatures. The trained models are then applied to screen for high-performance magnetic compounds.}
\label{fig:workflow}
\end{figure}

\section*{Results}
\subsection*{NEMAD: A Comprehensive Database of Magnetic Materials }

By applying state-of-the-art LLMs to scholarly \textit{experimental} articles published in Elsevier and American Physical Society (APS) journals, see details in the Methods section, we have constructed the NEMAD database to provide complete and accurate information on magnetic materials and their corresponding properties. The database includes 67,573 magnetic materials, uniquely defined
by chemical composition, structural detail, and magnetic properties. Each entry contains fifteen features listed in Table \ref{tab:Table1}. The chemical composition of each magnetic material was collected initially as string data, but was later converted into a numerical format by feature engineering techniques detailed in Methods, making it easy for training machine learning models.

\begin{table}[ht]
\centering
\caption{Description of Features in the NEMAD Database}
\label{tab:Table1}
\begin{tabular}{@{}llll@{}}
\toprule
\multirow{1}{*}{Column Name} & Type & Unit & Description \\
 & & & \\
\midrule
Material Chemical Composition & String &  & Chemical composition of the magnetic compound \\
Curie & Numeric & K & Curie temperature value of compound \\
N\'eel & Numeric & K & N\'eel temperature value of compound \\
Curie Weiss & Numeric & K & Curie Weiss temperature value of compound \\
Crystal Structure & String &  & Crystal structure of compound \\
Lattice Structure & String &  & Lattice structure of compound \\
Lattice Parameter & String & & Lattice parameter of compound \\
Space Group & String &  & Space group of compound \\
Coercivity & Numeric & Oe & Coercivity value of compound \\
Magnetization & Numeric & $\mathrm{AM}^{-1}$ & Magnetization value of compound \\
Magnetic Moment & Numeric & $\mu$B & Magnetic moment value of compound \\
Remanence & Numeric & $\mathrm{AM}^{-1}$ & Remanence value of compound \\
Susceptibility & Numeric &  & Susceptibility value of compound \\
DOI & String &  & Source of the entire collected information \\
Experimental & Boolean &  & Experimental status (Yes/No) of the article \\
\bottomrule
\end{tabular}
\end{table}

NEMAD includes both ferromagnetic and antiferromagnetic materials identified  by their temperature record. Around 68\% of the records in the database are ferromagnets. Records with solely N\'eel temperature classified as antiferromagnets make up about 30\% of the database. Only a tiny percentage (about 2\%) of materials have both Curie and N\'eel temperatures due to their complicated phase diagrams. The Curie-Weiss temperature is another feature that contains important information for the magnetic properties. Discrepancy between the Curie-Weiss temperature and transition temperature usually indicates competing interactions in the material. The distributions of Curie and N\'eel temperatures throughout the database are shown in Figure 2. It is noteworthy that almost 22\% of the compounds have Curie temperatures higher than 600K, suggesting broad temperature distributions and the possibility of using this database to search high Curie temperature materials.

\begin{figure}[ht]
\centering
\includegraphics[width=\linewidth]{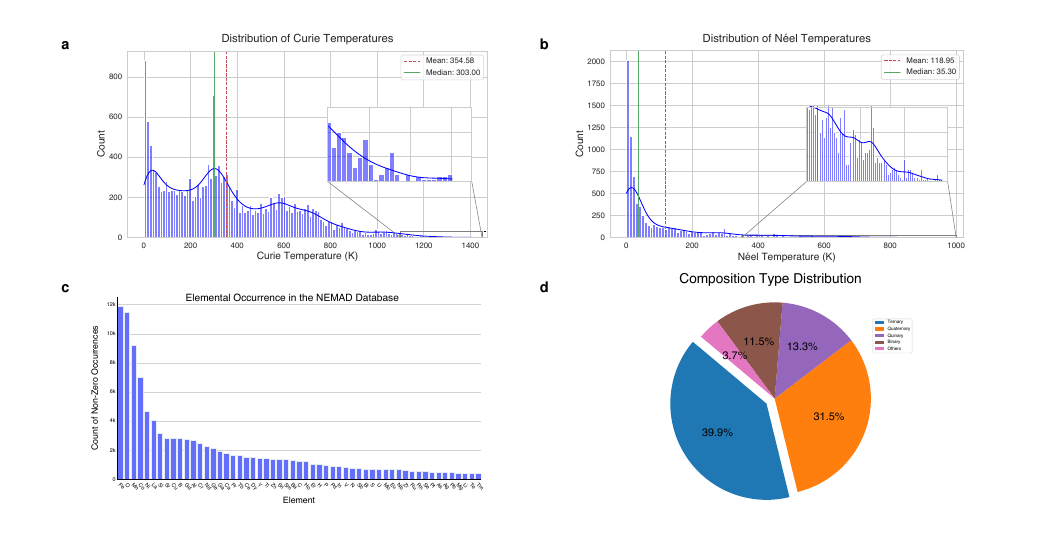}
\caption{\textbf{Comprehensive analysis of the NEMAD database.} Histogram display the distribution and frequencies of the (\textbf{a}) Curie  and (\textbf{b}) N\'eel  temperatures across the database. (\textbf{c}) Bar chart represents the frequency of each element present in the database, highlighting the most common elements in magnetic materials. Only elements with a frequency greater than 300 are included. (\textbf{d}) Pie chart categorizes materials into binary, ternary, quaternary, quinary, and higher-order compounds, depicting their relative proportions within the database.}
\label{fig:2}
\end{figure}

In addition, the database comprises structural details of the magnetic compounds such as crystal structure, lattice structure, lattice parameters, and space groups. This structural data is necessary for creating advanced machine learning models like graph neural networks. It also provides several magnetic property values in the relevant units, including coercivity, magnetization, magnetic moment, remanence, and susceptibility. These features are useful in data-driven discovery regarding high-performance magnetic materials, particularly permanent magnets.

The elemental composition of the NEMAD database is extensive, including 84 different elements. The frequencies of these elements in the database are summarized in Figure 2c. Materials containing elements with high Curie temperatures, such as iron (Fe), cobalt (Co), and nickel (Ni), are numerous. Furthermore, some other elements like Mn, La, Sr, Cu, B, Al, Cr, and Ce are also present in appreciable frequencies, which are necessary for the construction of high-performance magnetic materials. Noticeably, this database contains a huge number of magnetic compounds without rare earth elements. It is potentially useful for discovering permanent magnets without rare-earth elements. Figure 2d displays the distribution of compound types in our database. Approximately 40\% are ternary compounds  (composed of three distinct elements). The remaining compounds are predominantly quaternary, quinary, and binary.   

To validate the quality of the NEMAD database, we randomly selected 5,015 records and used a another large language model (Google Gemini 2.5) to independently evaluate each extracted field against the corresponding original article. The model assessed each entry based on whether the extracted value matched, was absent, or incorrectly assigned. The overall accuracy of the database is high, with a median accuracy of 94\% across all fields. A small number of manually reviewed samples further confirmed the consistency of the validation model. Full validation details and field-level performance metrics are provided in the Supplementary Information.

\subsection*{Machine Learning Model for Classifying Materials}

To classify the materials into non-magnet (NM), ferromagnet (FM), and anti-ferromagnet (AFM) categories, we trained a random forest classifier model based on the NEMAD database; see details in the Methods section. Random Forest was chosen due to its successful application in materials informatics literature, especially on small to medium sized datasets, and its robustness in handling heterogeneous, tabular data. Non-magnetic materials used to train the model are taken from the Materials Project. The classification was performed using features derived from chemical compositions, as described in detail in the Feature Engineering section.  A consistent accuracy of 0.89 on validation and 0.90 on testing sets underpins the strong generalization of our model toward unseen data. We provide detailed performance metrics for every class across the training, validation, and testing datasets in Table~\ref{tab:Table2}. Precision, recall, and F1-score for all classes on the validation and testing sets are consistent, demonstrating that the model is not biased toward any specific subset of data. The overall performance of the model is slightly better in the training set than in the testing set, indicating minor overfitting, but the model still remains robust. The AFM class has slightly lower performance metrics compared to other classes, which may be attributed to the smaller number of antiferromagnetic records in our training dataset. This conclusion is supported by the consistent underperformance of the AFM class across the validation and testing sets.

\begin{table}[ht]
\centering
\caption{Evaluation metrics used to assess the performance and effectiveness of classification models.}
\label{tab:Table2}
\begin{tabular}{@{}lllllll@{}}
\toprule
Model & Dataset & Class & Precision & Recall & F1-Score & Accuracy \\
\midrule
\multirow{9}{*}{Random Forest Classifier} 
& \multirow{3}{*}{Training (60\%)} & FM  & 1.00 & 0.99 & 1.00 & \multirow{3}{*}{1.00} \\
&                                  & AFM & 0.99 & 0.99 & 0.99 & \\
&                                  & NM  & 1.00 & 1.00 & 1.00 & \\
\cmidrule{2-7}
& \multirow{3}{*}{Validation (20\%)} & FM  & 0.90 & 0.91 & 0.91 & \multirow{3}{*}{0.89} \\
&                                    & AFM & 0.79 & 0.73 & 0.76 & \\
&                                    & NM  & 0.93 & 0.96 & 0.95 & \\
\cmidrule{2-7}
& \multirow{3}{*}{Testing (20\%)} & FM  & 0.91 & 0.91 & 0.91 & \multirow{3}{*}{0.90} \\
&                                 & AFM & 0.80 & 0.76 & 0.78 & \\
&                                 & NM  & 0.94 & 0.97 & 0.95 & \\

\midrule
\multirow{9}{*}{XGB Classifier}
& \multirow{3}{*}{Training (60\%)} & FM  & 1.00 & 0.99 & 1.00 & \multirow{3}{*}{1.00} \\
&                                  & AFM & 0.99 & 0.99 & 0.99 & \\
&                                  & NM  & 1.00 & 1.00 & 1.00 & \\
\cmidrule{2-7}
& \multirow{3}{*}{Validation (20\%)} & FM  & 0.91 & 0.92 & 0.91 & \multirow{3}{*}{0.90} \\
&                                    & AFM & 0.79 & 0.75 & 0.77 & \\
&                                    & NM  & 0.96 & 0.97 & 0.97 & \\
\cmidrule{2-7}
& \multirow{3}{*}{Testing (20\%)} & FM  & 0.91 & 0.92 & 0.91 & \multirow{3}{*}{0.91} \\
&                                 & AFM & 0.81 & 0.78 & 0.80 & \\
&                                 & NM  & 0.96 & 0.97 & 0.96 & \\

\bottomrule
\end{tabular}
\end{table}

In addition to the random forest model, we trained an XGB classifier on the same dataset using the same feature set. This model was selected for its track record in structured scientific data and its ability to capture non-linear relationships through gradient boosting. The XGB classifier model demonstrated nearly identical overall performance, with an accuracy of 0.90 on the validation set and 0.91 on the testing set. While there are slight differences in precision, recall, and F1-score for specific classes (FM, AFM, NM), the performance metrics remain highly consistent between the two models. This similarity further confirms the robustness and generalizability of our classification approach. The AFM class again showed slightly reduced precision and recall, consistent with the limited number of AFM training samples.

Figure 3a, 3b, 3d, and 3e present the confusion matrices, which further illustrate the classification performance across the validation and testing datasets.
In these matrices, the diagonal elements represent the number of correctly classified samples for each class, while the off-diagonal elements indicate misclassifications. Both classifiers show high accuracy, as most values are concentrated along the diagonal, with relatively few misclassifications between FM, AFM, and NM classes.

\begin{figure}[ht]
\centering
\includegraphics[width=\linewidth]{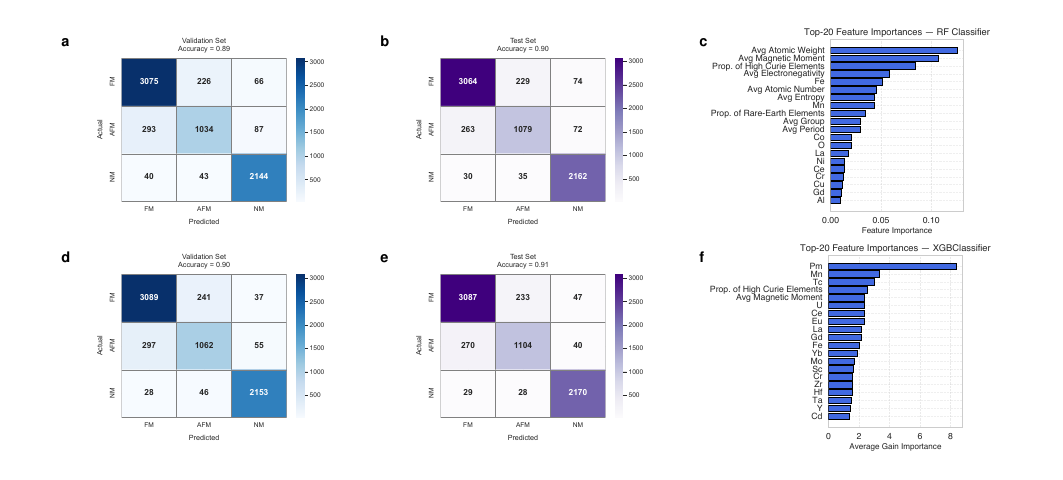}
\caption{\textbf{Performance and analysis of classification models for magnetic materials.} Top row (\textbf{a-c}): Results from the Random Forest (RF) classifier. Bottom row (\textbf{d-f}): Results from the eXtreme Gradient Boosting (XGBoost) classifier. Confusion matrices illustrate the performance on validation sets (\textbf{a,d}) and test sets (\textbf{b,e}), showing the number of true positive, true negative, false positive, and false negative predictions. The feature importance plots (\textbf{c,f}) rank the significance of various features used in each classification model. Feature importance scores are calculated as  mean decrease in gini impurity for the RF classifier and as the average gain of splits for the XGB classifier.}
\label{fig:3}
\end{figure}

Existing classification models in literature usually classify NM, FM, and AFM using two-step approach; the first step is to classify only non-magnetic and magnetic compounds, and only magnetic materials are passed into the second step to classify FM or AFM \cite{lu2022fly,long2021accelerating}. Our model achieves one-step classification successfully. Besides, almost all of them are trained on the DFT-generated dataset. In contrast, our model was trained on dataset of first hand experimental reports. It shows that only input of chemical composition is enough to classify the magnetic materials with appreciable accuracy.

Understanding which features contribute most significantly to classification decisions is essential for interpreting model behavior and gaining insight into the underlying feature-property relationships in the dataset. For both the RF classifier and XGB classifiers, we evaluated feature importance to identify the most influential descriptors used during training.
Figure 3c and 3f present the top 20 ranked features for the RF and XGB classifier models, respectively. For the RF classifier, feature importance was computed based on the mean decrease in impurity (MDI), which reflects the cumulative reduction in Gini impurity achieved by each feature across all decision trees in the ensemble. The most impactful features include the average atomic weight, average atomic magnetic moment, the proportion of high curie temperature elements, and average electronegativity. 
For the XGB classifier, we measured feature importance using the average gain. This metric shows how much each feature improves the model’s predictions when it is used to split the data.

\subsection*{Regression Model for Predicting Curie Temperature of Ferromagnetic Compounds}
In this section, we trained and compared different regression models, including Random Forest (RF) regressor, Extreme Gradient Boosting (XGBoost), and Ensemble Neural Network (ENN), toward accurate prediction of the Curie temperature for a ferromagnetic compound. These models were selected based on their complementary strengths and established success in prior materials informatics research. Tree-based models such as RF and XGBoost are known for their robustness, interpretability, and ability to handle tabular data with complex feature interactions, while neural networks are well-suited for capturing non-linear patterns in the data. These models were trained on two different datasets: one with all the unique experimentally verified ferromagnetic records in the NEMAD database, and another with a balanced dataset. The original dataset exhibits a highly skewed distribution of Curie temperatures, with a significant concentration of materials in the low-temperature regime. This imbalance can lead machine learning models to be biased toward more frequently occurring temperature ranges, degrading their generalization performance for higher-temperature compounds. To address this, we employed a stratified undersampling technique to construct a balanced dataset that provides uniform coverage across the entire Curie temperature range. Specifically, we partitioned the Curie temperature values into discrete bins and applied stratified sampling to ensure equal representation of samples from each bin. This approach enhances the diversity and distributional balance of the training data, helping to mitigate model bias toward overrepresented regions and improving performance across the full temperature spectrum.

The bin-based stratified undersampling was integrated into a cross-validated ensemble training procedure, in which individual models were independently trained on balanced subsets from each fold. A final ensemble consisting of 30 models was trained on the full stratified training set to generate robust predictions and estimate uncertainty. 
For details on dataset distribution and the full model development pipeline, see the Methods section.

Model performance was measured with standard metrics, including R-squared (R$^2$), mean absolute error (MAE), and root mean squared error (RMSE). 
Table~\ref{tab:Table3} provides detailed performance metrics of these models for the test set of each dataset. For the original dataset, the XGBoost model achieved an R$^2$ value of 0.84, MAE of 62~K, and RMSE of 107~K, compared to RF (R$^2$ = 0.82, MAE = 71~K, RMSE = 115~K) and ENN (R$^2$ = 0.84, MAE = 60~K, RMSE = 110~K). Similarly, for the balanced dataset, the XGBoost model reached an R$^2$ value of 0.87, MAE of 56~K, and RMSE of 97~K, while RF and ENN attained (R$^2$ = 0.85, MAE = 65~K, RMSE = 104~K) and (R$^2$ = 0.85, MAE = 56~K, RMSE = 103~K), respectively. While the differences are not substantial enough to claim superiority, XGBoost consistently performs at top across multiple metrics and datasets. Notably, all models performed better using the balanced dataset; in particular, performance improved most for the XGBoost model. On this balanced dataset, this resulted in a 4\% improvement in R$^2$ for the XGBoost model and 4\% and 1\% improvement for the RF and ENN models, respectively. Slightly higher performance on the balanced dataset indicates that models are better at capturing the relationships within higher temperature ranges. This could be an improvement due to the increased relative representation of the higher temperature data.

\begin{table}[htbp!]
\centering
\caption{Evaluation metrics used to assess the performance and effectiveness of various regression models trained on two datasets for predicting Curie temperature.}
\label{tab:Table3}
\begin{tabular}{@{}lllll@{}}
\toprule
\multirow{1}{*}{Model} & Dataset & R-squared & MAE & RMSE \\
 &  &(Test)  &(Test)  &(Test)   \\
\midrule
\multirow{2}{*}{Random Forest} & Original  & 0.82 & 71K & 115K\\
& Balanced (Undersampled)& 0.85 & 65K & 104K\\
\multirow{2}{*}{Ensembled Neural Network} & Original  & 0.84 & 60K & 110K\\
& Balanced (Undersampled)& 0.85 & 56K & 103K \\
\multirow{2}{*}{XGBoost} & Original  & 0.84 & 62K & 107K\\
&Balanced (Undersampled)& 0.87 & 56K & 97K \\
\bottomrule
\end{tabular}

\end{table}

The plots between the predicted and actual Curie temperature for the test set of the balanced dataset are shown in Figures 4d, 4e, and 4f. In addition, the plots include prediction error bars (±1$\sigma$) derived from the model ensembles, providing visual estimates of confidence intervals for each prediction. This model based on the balanced dataset has better performance than that on the original dataset, as shown in Figures 4a, 4b, and 4c. In the region where the Curie temperature is greater than 500~K in the original dataset, all three models underestimated the Curie temperature. This suggests the necessity of expanding the database by adding higher Curie temperature data points in the future.

To further assess the model’s predictive accuracy, we conducted an error analysis using the absolute
error plots. Figures 4g, 4h, and 4i  presents these plots for all three models across balanced dataset. We fitted an exponential curve to each
error distribution to characterize the error pattern. In XGBoost and ENN, 70\% of the test data have an
absolute error of less than 50K, whereas in RF model, this proportion is 62\%.

\begin{figure}[ht]
\centering
\includegraphics[width=\linewidth]{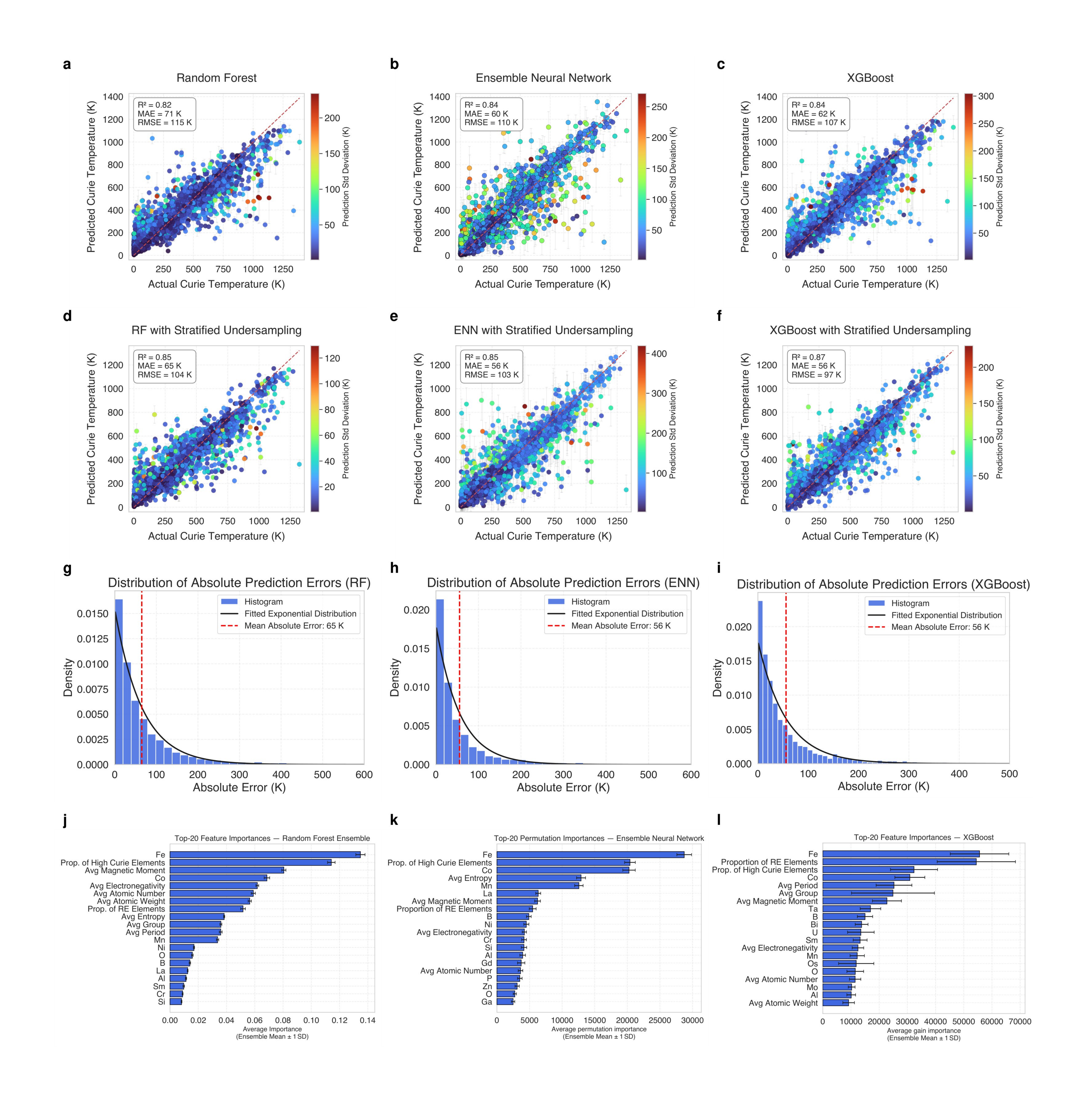}
\caption{\textbf{Evaluation of Curie temperature prediction models on original and stratified-balanced datasets.}
(\textbf{a–c}) Predicted versus actual Curie temperatures for the test set of the original dataset using three models: Random Forest (\textbf{a}), Ensemble Neural Network (\textbf{b}), and eXtreme Gradient Boosting (XGBoost) (\textbf{c}). 
(\textbf{d–f}) Corresponding predictions on a balanced dataset created using stratified undersampling. To construct this dataset, the Curie temperature range was divided into bins, and samples from the overrepresented low-temperature region were undersampled to create a more uniform temperature distribution. 
Each plot reports the coefficient of determination (R$^2$), mean absolute error (MAE), and root mean squared error (RMSE), and includes confidence intervals based on ensemble standard deviation.
(\textbf{g–i}) Absolute error distributions and fitted exponential curves for the balanced dataset across the three models: Random Forest (\textbf{g}), Ensemble Neural Network (\textbf{h}), and XGBoost (\textbf{i}). 
(\textbf{j–l}) Feature importance plots showing the top 20 most influential features for models trained on the balanced dataset: Random Forest (\textbf{j}), Ensemble Neural Network (\textbf{k}), and XGBoost (\textbf{l}).
All models were trained on features derived from the chemical composition of materials in the NEMAD database. The figure provides a comparative assessment of model performance, uncertainty, and feature relevance under both original and balanced training conditions.}
\label{fig:4}
\end{figure}

To better understand the contribution of each feature in predicting curie temperature, we computed model-specific feature importances using methods appropriate for each algorithm. For the RF model, feature importance was calculated as the average reduction in impurity across trees in the ensemble. For the XGBoost model, we used the average gain metric, which quantifies the improvement in the loss function brought by each feature across all decision trees. For the ENN, permutation importance was applied, which measures the increase in prediction error when the values of a given feature are randomly shuffled. In all cases, feature importances were averaged across 30 independently trained models, and standard deviations were reported to reflect the variability across the ensemble.
Figures 4j, 4k, and 4l display the top 20 features ranked by importance for each model. Notably, features such as the proportion of Fe and Co atoms, the proportion of high-Curie-temperature elements, the average magnetic moment, and the presence of rare-earth elements consistently appeared among the most influential predictors across all models.

\subsection*{Regression Model for Predicting N\'eel Temperature of Antiferromagnetic Compounds}

Similar methods can be developed to predict N\'eel temperatures of antiferromagnetic compounds. Details of procedures for data cleaning and model building are explained in the Methods section. We trained and evaluated three distinct models: RF regressor, ENN, and XGBoost. Among these, ENN and XGBoost demonstrated consistently strong performance, while RF showed slightly lower accuracy in error metrics. XGBoost achieved an R$^{2}$ value of 0.83, followed by ENN with an R$^{2}$ of 0.80. The RF model yielded a comparable R$^{2}$ value of 0.81.

The MAE and RMSE further illustrate the performance of the models. The XGBoost model had an MAE of 38~K and RMSE of 72~K, while ENN achieved an MAE of 38~K and RMSE of 76~K. The RF model showed slightly higher error rates, with an MAE of 43~K and RMSE of 76~K. Although the differences in performance are modest, XGBoost and ENN show consistently competitive results across both metrics.

Figure 5 presents the predicted versus actual N\'eel temperature plots, along with absolute error distributions and feature importance plot for the RF, ENN, and XGBoost models. Similar to the case of Curie temperature prediction, the deviation in predicted values is more pronounced at higher temperatures, likely due to the limited availability of high-N\'eel-temperature training samples. The absolute error plots demonstrate that approximately 64\% of data points in the test set have an absolute error below 25~K for the XGBoost model, whereas in the ENN model, this proportion is 63\%. For the Random Forest model, around 60\% of predictions fall within this error threshold.

\begin{figure}[ht]
\centering
\includegraphics[width=\linewidth]{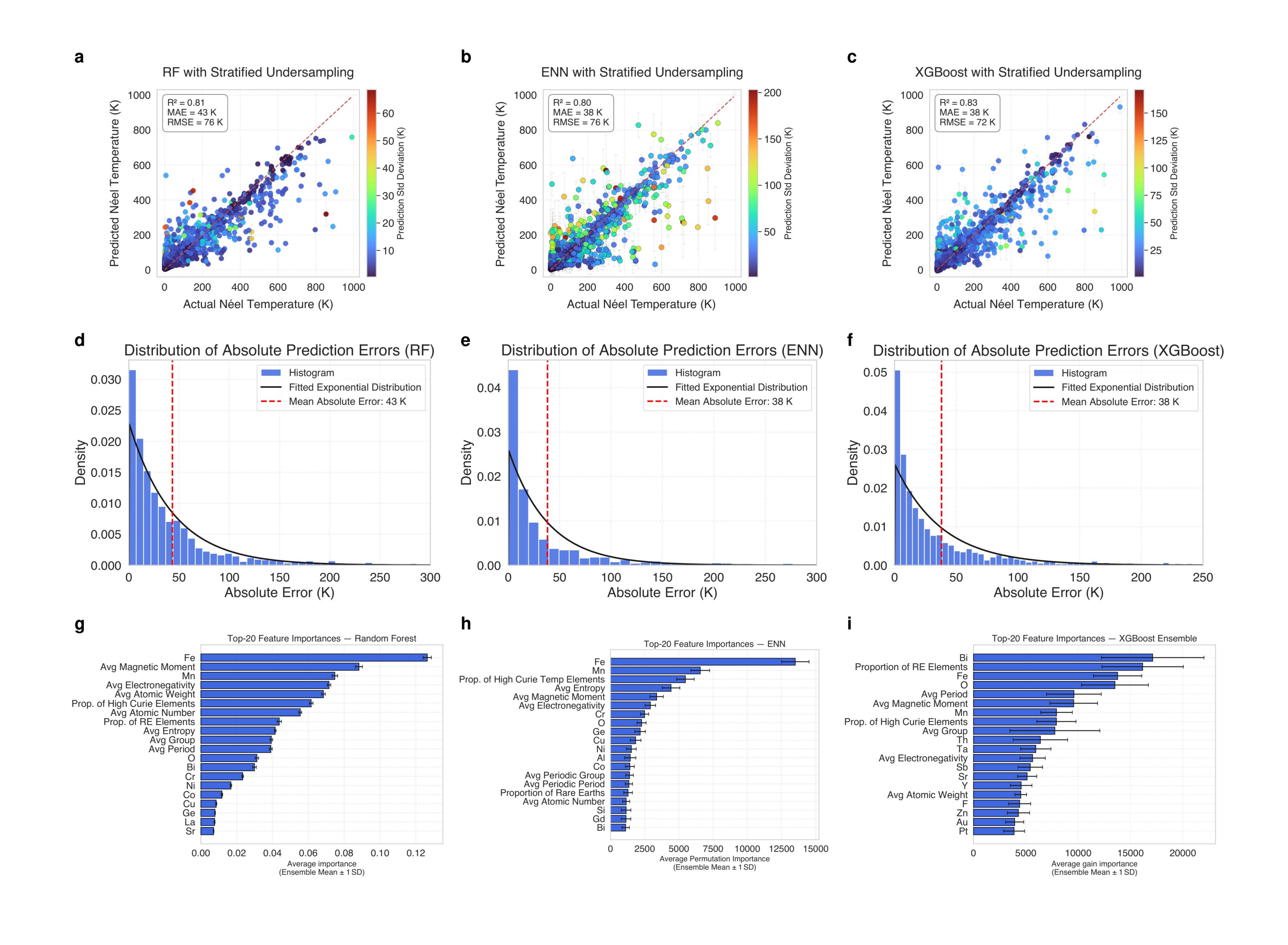}
\caption{\textbf{Evaluation of machine learning models for N\'eel temperature prediction using stratified undersampling.} 
Top row (\textbf{a–c}): Predicted versus actual N\'eel temperatures on the test set using three models-Random Forest (\textbf{a}), Ensemble Neural Network (\textbf{b}), and eXtreme Gradient Boosting (\textbf{c})-trained on a balanced dataset created via stratified undersampling. Each plot displays the coefficient of determination (R$^2$), mean absolute error (MAE), and root mean squared error (RMSE). The predictions are color-coded by the standard deviation across ensemble predictions, and error bars reflect the confidence interval associated with each prediction. 
Middle row (\textbf{d–f}): Distribution of absolute prediction errors on the test set for the same models, with an exponential fit overlaid and the mean absolute error marked by a dashed red line. 
Bottom row (\textbf{g–i}): Top-20 feature importance plots for each model, showing the most influential descriptors used in predicting the N\'eel temperature.}
\label{fig:5}
\end{figure}

Each scatter plot includes both error bars and color-coded points showing the standard deviation of predictions across ensemble models. This provides a visual indication of prediction uncertainty for all three models. Compared to RF and XGBoost, the ENN shows noticeably larger confidence intervals. This reflects higher uncertainty in its predictions. The ENN is a highly nonlinear model, which tends to be more sensitive to small changes in data. In our case, we used stratified undersampling to train 30 separate ENN models, each on a different balanced subset of the data. The undersampling was applied only to the low N\'eel temperature region, meaning that each model was trained on a different subset of low-temperature samples. This introduces variation in model behavior and results in higher standard deviation across predictions. These wider confidence intervals suggest that the ENN predictions are less reliable for compounds with lower N\'eel temperatures. In contrast, the RF and XGBoost models show narrower confidence intervals, indicating more stable and consistent predictions across ensemble members.

Feature importance plots show the top 20 most influential features in predicting the N\'eel temperature of antiferromagnetic compounds. All three models identified the proportion of iron atom (Fe) in the chemical composition as an important feature. A significant number of transition metal oxides, including those of iron, manganese, cobalt, and nickel, exhibit antiferromagnetic behavior. This arises from superexchange interactions between magnetic ions mediated by oxygen anions, which favor anti-parallel spin alignment, leading to antiferromagnetic ordering~\cite{coey2021magnetic, oles2003magnetic}. The consistent appearance of oxygen among the top-ranked features in all models further supports that the models have learned physically meaningful patterns.

\subsection*{Screening High Performance Magnetic Material from External Databases}

Building on the successful training and evaluation of our machine learning models, we now apply them to a separate, previously unseen dataset to identify materials with targeted properties. To this end, the Materials Project database was selected, from where marked magnetic materials with features like chemical composition, magnetic phase, and stability were downloaded using an API key. First, we removed materials which are already included in our database. Although every material from the Materials Project has a label of its magnetic state, we still implemented our classification model on the dataset to reconfirm the magnetic state, i.e., ferromagnetic, antiferromagnetic, or non-magnetic, of each compound. We listed 602 ferromagnetic and 237 antiferromagnetic compounds having the consensus between our classification results and the Materials Project database. We then used our regression models to predict the Curie and N\'eel temperature of all the above-listed ferromagnetic and antiferromagnetic compounds. 

In addition to the Materials Project, we also applied our models to another source of magnetic materials: a curated set of thermodynamically stable Heusler and half-Heusler compounds reported in a high-throughput DFT study~\cite{jiang2022high}. These compounds were first checked to remove any duplicates already present in our database. Then, we applied the same classification and regression models to determine their magnetic state and predict their Curie or N\'eel temperatures.

Table~\ref{tab:Table4} presents a selection of 25 newly identified ferromagnetic compounds with predicted Curie temperatures exceeding 500K in at least two models, along with 13 antiferromagnetic compounds with predicted Néel temperatures above 100K based on the XGBoost model. These candidates were drawn from both the Materials Project and the DFT-verified Heusler datasets.
In total, our screening identified 32 high-probability magnetic compounds, of which 7 have since been found in the literature with experimentally reported Curie temperatures, validating our model predictions. The remaining 25 represent candidates not previously reported with experimental magnetic ordering. 
Future experimental verification of these predictions is strongly encouraged.

\begin{longtable}{@{}lcccccc@{}}
\caption{List of candidate ferromagnetic and antiferromagnetic compounds with high predicted Curie and N\'eel temperatures, as estimated by our three best-performing machine learning models. Compounds marked with \textsuperscript{*} have experimental ordering temperatures reported in the literature (see Comment column), while all other candidates remain to be experimentally verified.}
\label{tab:Table4} \\

\toprule
Composition & Type & Crystal System& XGBoost (T$_C$) & ENN (T$_C$)& RF (T$_C$)& Comment  \\
\midrule
\endhead

\midrule
\multicolumn{5}{r}{Continued on next page} \\
\endfoot

\bottomrule
\endlastfoot
VFeCoGe & FM &Cubic& $500.0 \pm 57.2 \text{K}$ & $256.0 \pm 142.4 \text{K}$ & $556.0 \pm 19.0 \text{K}$& \\

LiNbFeO4 & FM& Tetragonal& $511.7 \pm 78.0 \text{K}$ & $587.9 \pm 202.2 \text{K}$ & $424.1 \pm 21.4 \text{K}$ \\

Fe2NiP & FM &Tetragonal & $582.1 \pm 36.1 \text{K}$ & $381.7 \pm 83.1 \text{K}$ & $583.1 \pm 10.8 \text{K}$ \\

GaFe2Co4Si & FM &Trigonal& $1004.9 \pm 35.7 \text{K}$ & $1009.7 \pm 60.8 \text{K}$ & $978.8 \pm 16.8 \text{K}$ \\

Fe3PdN & FM &Cubic& $510.7 \pm 54.7 \text{K}$ & $600.1 \pm 174.9 \text{K}$ & $478.1 \pm 27.3 \text{K}$ \\

AlFe3H & FM &Cubic& $569.0 \pm 61.8 \text{K}$ & $711.0 \pm 101.9 \text{K}$ & $550.5 \pm 36.3 \text{K}$ \\

Fe3Rh & FM & Cubic& $645.6 \pm 39.6 \text{K}$ & $878.6 \pm 125.9 \text{K}$ & $572.8 \pm 20.9 \text{K}$ \\

Hf(GaFe)6 & FM & Orthorhombic&$527.0 \pm 31.4 \text{K}$ & $500.0 \pm 108.3 \text{K}$ & $466.1 \pm 12.8 \text{K}$ \\
Fe3Co3Si2 &FM &Trigonal &$945.3 \pm 41.5 \text{K}$ & $999.0 \pm 68.8 \text{K}$ & $900.9 \pm 20.7 \text{K}$ \\

AlFe3C & FM &Cubic& $566.8 \pm 58.3 \text{K}$ & $388.3 \pm 188.1 \text{K}$ & $502.3 \pm 40.0 \text{K}$ \\

Ga3Fe4Co8Si & FM &Trigonal &$998.7 \pm 36.3 \text{K}$ & $1046.2 \pm 72.0 \text{K}$ & $938.9 \pm 13.9 \text{K}$ \\

Zr(GaFe)6 & FM &Orthorhombic &$500.6 \pm 34.1 \text{K}$ & $520.6 \pm 168.7 \text{K}$ & $482.9 \pm 14.1 \text{K}$ \\

Mn2GaCo4Ge & FM & Trigonal&$658.0 \pm 43.6 \text{K}$ & $613.9 \pm 148.3 \text{K}$ & $455.8 \pm 38.2 \text{K}$ \\

Mn4Cr(Co2Ge)5 & FM &Trigonal &$557.3 \pm 80.4 \text{K}$ & $661.8 \pm 183.5 \text{K}$ & $444.9 \pm 48.6 \text{K}$ \\

Fe3RhN & FM &Cubic &$524.0 \pm 54.2 \text{K}$ & $677.6 \pm 170.1 \text{K}$ & $472.3 \pm 26.6 \text{K}$ \\

Fe4Li3SbO8 & FM & Monoclinic&$712.6 \pm 53.5 \text{K}$ & $691.0 \pm 189.7 \text{K}$ & $630.9 \pm 16.2 \text{K}$ \\

MnZn3(CrSe2)8 & FM &Trigonal& $788.3 \pm 117.4 \text{K}$ & $982.5 \pm 115.1 \text{K}$ & $560.5 \pm 75.4 \text{K}$ \\

VFe2BO5 & FM &Orthorhombic& $552.7 \pm 70.4 \text{K}$ & $288.8 \pm 144.9 \text{K}$ & $514.8 \pm 18.2 \text{K}$ \\

GaFeNiCo & FM &Cubic& $923.2 \pm 58.8 \text{K}$ & $848.4 \pm 151.5 \text{K}$ & $872.4 \pm 43.1 \text{K}$ \\

RhCoGaFe & FM & Cubic&$660.8 \pm 60.4 \text{K}$ & $660.8 \pm 60.4 \text{K}$ & $609.0 \pm 18.1 \text{K}$ \\

FeGaRuCo & FM &Cubic &$644.2 \pm 56.2 \text{K}$ & $617.5 \pm 137.4 \text{K}$ & $595.2 \pm 20.3 \text{K}$ \\

GeFeNiCo & FM &Cubic &$740.7 \pm 74.0 \text{K}$ & $856.1 \pm 187.1 \text{K}$ & $633.8 \pm 28.6 \text{K}$ \\

TcCoGeFe & FM &Cubic &$637.0 \pm 56.1 \text{K}$ & $692.3 \pm 161.1 \text{K}$ & $567.5 \pm 19.0 \text{K}$ \\

Fe2RhGa & FM &Cubic &$569.7 \pm 43.2 \text{K}$ & $595.9 \pm 147.6 \text{K}$ & $483.8 \pm 15.1 \text{K}$ \\

Co2NiGe & FM & Tetragonal&$664.4 \pm 70.8 \text{K}$ & $846.1 \pm 184.2 \text{K}$ & $483.6 \pm 24.0 \text{K}$ \\

Fe3PtN \small ${}^*$ & FM &Cubic &$614.4 \pm 43.1 \text{K}$ & $706.0 \pm 144.5 \text{K}$ & $445.2 \pm 17.3 \text{K}$ &  \small ${}^*$ 642K \cite{wiener1955structure}\\
%experimental 642K 10.1007/BF03377510

Y(GaFe)6 \small ${}^*$ & FM & Orthorhombic&$463.7 \pm 20.1 \text{K}$ & $431.7 \pm 52.1 \text{K}$ & $437.3 \pm 17.8 \text{K}$& \small ${}^*$ 500 K \cite{weitzer1990magnetism} \\ % 

Co2MnSb \small ${}^*$  & FM &Cubic &$590.7 \pm 34.3 \text{K}$ & $592.9 \pm 110.2 \text{K}$ & $533.1 \pm 41.9 \text{K}$ & \small ${}^*$  $600\pm 10 \text{K}$ \cite{webster1971magnetic}\\

Fe2CuGa \small ${}^*$ & FM &Tetragonal &690.9 ± 73.3 K & 716.2 ± 173.6 K & 565.9 ± 16.8 K & \small ${}^*$ 798 K \cite{gasi201357}\\

GeCrCoFe\small ${}^*$ & FM & Cubic&$634.1 \pm 85.7 \text{K}$ & $416.4 \pm 208.5 \text{K}$ & $623.9 \pm 30.1 \text{K}$ & \small ${}^*$866 K  \cite{enamullah2015electronic}\\

NbAlCo2 \small ${}^*$& FM &Cubic& $329.2 \pm 62.2 \text{K}$ & $386.6 \pm 175.2 \text{K}$ & $382.8 \pm 15.7 \text{K}$ &\small ${}^*$ 383~K\cite{buschow1981magnetic} \\

Fe2RuGe\small ${}^*$ & FM &Cubic& $573.2 \pm 27.4 \text{K}$ & $571.4 \pm 121.4 \text{K}$ & $448.6 \pm 25.3 \text{K}$ & \small ${}^*$860~K  \cite{chakraborty2023observation}\\

Sr2FeBrO3 & AFM &Tetragonal &$284.8 \pm 30.6 \text{K}$ & $119.1 \pm 65.8 \text{K}$ & $249.5 \pm 6.8 \text{K}$ \\

Sr2FeClO3 & AFM & Tetragonal&$290.2 \pm 34.5 \text{K}$ & $157.6 \pm 77.0 \text{K}$ & $238.5 \pm 8.6 \text{K}$ \\

BaFeP(O2F)2 & AFM &Monoclinic &$224.1 \pm 40.9 \text{K}$ & $203.3 \pm 163.8 \text{K}$ & $128.0 \pm 9.7 \text{K}$ \\

K2NaFeO3 & AFM &Orthorhombic &$137.6 \pm 43.7 \text{K}$ & $81.0 \pm 101.0 \text{K}$ & $160.8 \pm 10.0 \text{K}$ \\

Li2FeO3 & AFM & Monoclinic&$138.3 \pm 31.1 \text{K}$ & $34.3 \pm 18.8 \text{K}$ & $138.3 \pm 11.4 \text{K}$ \\

Ba3Fe2Br2O5 & AFM &Cubic &$189.5 \pm 49.5 \text{K}$ & $54.0 \pm 52.4 \text{K}$ & $181.1 \pm 14.6 \text{K}$ \\

Cr3(FeO6)2 & AFM & Monoclinic&$101.0 \pm 32.9 \text{K}$ & $95.4 \pm 66.8 \text{K}$ & $69.5 \pm 6.9 \text{K}$ \\

LiFeAsO4 & AFM &Orthorhombic &$110.1 \pm 30.4 \text{K}$ & $32.9 \pm 11.3 \text{K}$ & $97.8 \pm 10.0 \text{K}$ \\

NaFeAsO4F & AFM & Monoclinic&$138.7 \pm 39.8 \text{K}$ & $56.3 \pm 52.1 \text{K}$ & $91.7 \pm 6.6 \text{K}$ \\

SrCrHO2 & AFM &Tetragonal &$105.4 \pm 30.4 \text{K}$ & $127.5 \pm 71.7 \text{K}$ & $85.5 \pm 6.0 \text{K}$ \\

IrFeSnHf & AFM &Cubic& $129.3 \pm 29.4 \text{K}$ & $160.8 \pm 132.0 \text{K}$ &  $148.6 \pm 9.9 \text{K}$ \\

HfSnRhFe & AFM &Cubic& $132.2 \pm 26.6 \text{K}$ & $113.6 \pm 81.5 \text{K}$ & $152.1 \pm 9.2 \text{K}$ \\

IrFeSnTi & AFM &Cubic &$157.1 \pm 32.6 \text{K}$ & $313.1 \pm 179.7 \text{K}$ & $168.6 \pm 8.6 \text{K}$ \\
\end{longtable}

\section*{Discussion}

It is interesting to note that our database only includes 30\% antiferromagnetic materials, far less than ferromagnets. One reason may be because of the presence of many ferromagnetic alloys in the database. The number of alloys is in principle unbounded and those containing magnetic elements such as iron are more likely to be ferromagnetic due to their itinerancy nature. On the other hand, there exist many antiferromagnets in transition metal oxides. Experimental reports for this community were published in journals outside of Elsevier and APS. It is thus important to expand the coverage of articles to include other publishers such as the Springer. A large database on magnetic materials is expected. 

A key product of this study is the development of a user-friendly website \href{http://www.nemad.org}{www.nemad.org} that hosts the NEMAD database. This website systematically organizes all the materials, allowing users to easily access detailed information about each material's properties. Users can explore the database and retrieve material data directly from the platform. We will continue to increase our database size. 

Our database also include the structural information of magnetic materials. This can be used in building a more powerful models, such as the graph neural network, in predicting magnetic properties. We also built a simple XGBoost machine learning model to incorporate these structure features like crystal system and space group (detailed included in the Supplementary Information). The XGBoost model achieved an R$^{2}$  value of 0.83, MAE of 52K, and a RMSE of 98K. The MAE and RMSE values are slightly improved compared to the model without structural information. 

Our LLM-based approach of extracting information and creating automated databases is versatile and can be applied to other areas of material science, such as superconducting, thermoelectric, photovoltaic, ferroelectric materials, etc. This method has the potential to transform how we gather and use scientific knowledge. The method used here to build the predictive models can also be adapted to predict other critical properties of magnetic materials, such as coercivity and saturation magnetization.

In summary, this study presents an effective method for discovering high-performance magnetic materials using large language models and machine learning techniques. A comprehensive database of magnetic materials was built. It includes not only the chemical and magnetic information, but also structural information. The machine learning models trained based on this database can be used to accurate prediction of new magnetic materials. These tools can be employed by researchers to develop next-generation magnetic materials through large-scale materials screening .

\section*{Methods}
\subsection*{Database Compilation}
Figure 6 illustrates the detailed workflow of the method used in this work. We began by compiling a comprehensive list of $100,000$ Digital Object Identifiers (DOIs) of scientific articles related to Magnetic Materials by searching relevant keywords (Ferromagnetic, Anti-ferromagnetic, Curie, and N\'eel, Coercivity, Magnetization, Magnetocrystalline Anisotropy, Remanence) from journal platforms maintained by Elsevier and the American Physical Society. We downloaded the article for all listed DOIs using the authenticated API request. These articles were primarily obtained in XML format, which often contains a hierarchical structure with nested elements.
We developed a custom XML parsing script to extract the full content of articles, including tabular data, from pre-downloaded XML-formatted documents. The script converts all unstructured textual and tabular information into plain text markdown format and then saves it as a CSV file with the following features.

GPTArticleExtractor\cite{zhang2024gptarticleextractor} method provides a way to make a comprehensive material property database by automatically extracting data from scientific articles. This method leverages the cutting edge Large Language Models like GPT-3.5 by incorporating the OpenAI API key. One can extract their desired information from articles by just modifying the prompts used in the workflow. So effective prompts are crucial for the accuracy and efficiency of this method.

Although the original workflow demonstrated strong performance when processing lengthy articles using GPT-3.5 and GPT-4 models, it exhibited several limitations. First, it was unable to extract information from tables, which often contain key material properties not explicitly mentioned in the body text. This omission increased the risk of missing essential data. Second, the prompts used in the previous workflow lacked depth and specificity. In several instances, the language models failed to interpret the prompt correctly, particularly when articles reported multiple compositions and their corresponding properties. As a result, some relevant compositions were either omitted or extracted in inconsistent formats, making downstream processing and integration into the database challenging.
Moreover, when chemical compositions were expressed in variable stoichiometric forms, such as Fe$_x$Co$_{1-x}$, the previous workflow often failed to extract the actual values of the variables. These values are crucial in materials science, and their omission renders the extracted composition scientifically meaningless. Another significant limitation was the workflow’s dependence solely on API-based access to XML-formatted articles. It was not designed to process Portable Document Format (PDF) articles, scanned documents, or older scientific handbooks from the 20th century that are typically stored as image-based files.

\begin{itemize}
    \item DOI
    \item Title
    \item Abstract
    \item Description (body part of the article)
\end{itemize}

\begin{figure}[ht]
\centering
\includegraphics[width=0.8\linewidth]{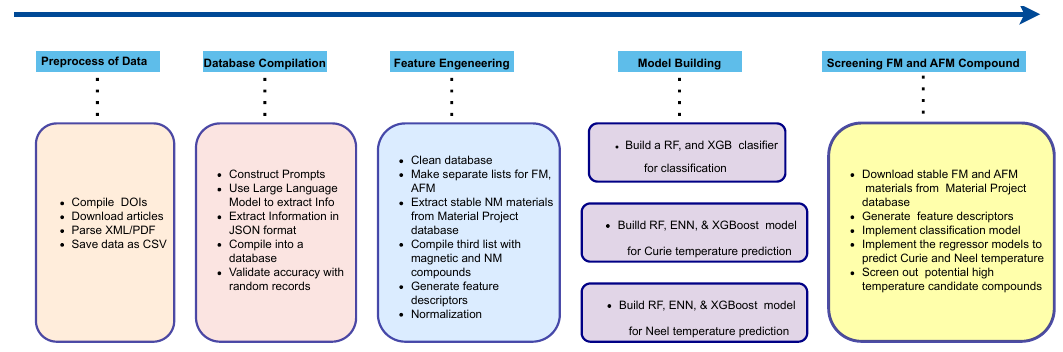}
\caption{\textbf{Detailed workflow for high-performance magnetic material discovery.} The workflow is divided into key stages, each represented by a block containing brief bullet points summarizing the methods and techniques employed. These blocks detail the sequential steps from  DOIs compilation to  materials screening, illustrating the comprehensive approach taken to screen high-performance magnetic materials.}
\label{fig:6}
\end{figure}

To overcome these limitations, we modified the previous GPTArticleExtractor workflow by leveraging the capabilities of the more powerful GPT-4o model. The enhanced pipeline supports XML articles accessed via API, as well as PDF documents, scanned image-based articles, and digitized historical handbooks. By expanding the input modalities and refining the prompt design, this upgraded workflow provides a more robust and inclusive data extraction process for constructing a comprehensive materials property database.

We listed the desired properties to extract from articles, including material chemical composition, magnetic phase transition temperatures (Curie, N\'eel, and Curie-Weiss), structural information (Crystal Structure, Lattice Structure, Lattice Parameters, and Space Group), and magnetic properties (Coercivity, Magnetization, Magnetic Moment, Remanence, and Susceptibility). We carefully reviewed literature from a range of publications to comprehend the trends and writing styles associated with these attributes to create effective prompts. For each property group, we designed a corresponding question prompt tailored to extract that specific type of information from the article. For instance, to extract material chemical composition, we formulated the prompt like this: ``Provide an exact chemical composition of the magnetic materials studied in the article. Some materials' compositions are expressed with variable proportions of elements, so include any values of such variables if mentioned." Similar prompts were constructed for structural information, magnetic transitions, and magnetic properties, forming a modular querying framework. 

Large Language Models are still constrained by input token limitations. It is less effective to use entire sections of a paper as input to these models, especially in the case of longer articles where non-relevant content may dilute the model's focus and exceed token limits. To address this, we designed a two-stage approach where question prompt was used to retrieve only the most relevant content from long-form articles. We first tokenized the articles by segmenting the text into chunks of 500 tokens each. Following that, these segments were examined in a vector space to determine word similarity using metrics like Euclidean distance. By comparing the vectorized text segments with particular question prompts, the model figured out the five most relevant segments using Facebook AI Similarity Search (FAISS) \cite{johnson2019billion}. This targeted input strategy allowed the model to focus on the most contextually relevant text, thereby improving response accuracy while reducing input token count and computing expenses.  
After identifying and collecting the most relevant text segments from the articles, the LLM was used to extract the final structured information using a summarization prompt. To ensure output consistency and format adherence, we used a one-shot prompting strategy in which the prompt also contained a single illustrative example of the expected output format. This technique enabled the generation of accurate and structured outputs. See the supplementary information for the full prompt templates.

Even with the use of a more powerful model and improved prompts, some important information was still missed. In particular, the model struggled with chemical compositions that had variable stoichiometry. This issue often occurred when an article contained many different compositions and discussed them throughout the entire article.  The tokenized chunks sometimes left out useful parts of the text, which led to incomplete data extraction. While the updated workflow was designed to manage token limitations in earlier models, recent advances in language models, such as GPT-4o, have increased input token capacity. As a result, it is now possible to process nearly the full content of an article. We found that removing non-informative section such as references allowed us to fit the majority of the relevant content within the model’s token limit. This adjustment further improved extraction accuracy, as it preserved the core technical sections containing material property data. We found that this comprehensive approach outperformed the previous technique, resulting in outputs that were more complete, accurate, and well-structured. This approach managed various material compositions and properties present in a single article by accurately creating list of jsons, where each entry includes all the information associated with a single chemical composition.
 Some outputs extracted by our model included only the chemical composition but lacked all other property information. Therefore, we established an inclusion criterion for the database: a material entry must contain at least one phase transition temperature or at least one magnetic property in addition to its composition.

To handle the PDF articles, we utilized a PDF parser \cite{wang2024mineruopensourcesolutionprecise,he2024opendatalab} to first convert the PDF into a structured markdown format. Although it is technically possible to pass the raw PDF directly to the language model, we observed that this often leads to suboptimal extraction performance. This is primarily because raw PDFs lack a consistent textual structure and require an intermediate parsing step to resolve layout complexities such as multi-column formatting, table structures, and chemical notations. During PDF to text conversion, crucial associations between material properties and compound names can become fragmented due to token chunking and the loss of layout context, which degrades extraction model accuracy. In contrast, the markdown format preserves hierarchical and semantic cues (e.g., headings, bullet points, tables) that significantly improve the model’s ability to extract and organize information reliably.
We then applied the same procedure to process all PDF articles and consolidated the extracted content into a single CSV file, where each row corresponds to a distinct article in markdown form.  This structured dataset, along with carefully designed prompts, was subsequently passed to the language model to generate the required information in a consistent and organized format.

Likewise, to extract information from scanned PDF-formatted scientific articles, we used the Gemini 2.0 Flash model to first convert them into a structured markdown format. This model employs advanced multimodal capabilities to process both image and text inputs. Since scanned PDFs often contain only rasterized page images without any embedded text layer, direct information extraction is not feasible. Therefore, the model first applies optical character recognition (OCR) to each page to recover the underlying textual content. Gemini 2.0 Flash leverages Google’s proprietary OCR and layout detection tools to accurately extract not only the raw text but also its structural context-such as headings, paragraph blocks, tables, and figure captions. Following OCR, the model reconstructs the document’s logical hierarchy and semantic structure, mapping it into markdown syntax with appropriate formatting cues (e.g., headings, bullet points, and table structures). This conversion is essential for improving the language model’s ability to identify and extract relevant scientific information consistently. While direct conversion is feasible for shorter scientific articles, it becomes impractical for longer documents such as handbooks due to the input token limitations of the language model. To address this, we first segmented each handbook into individual pages and processed each page separately through the model. The output from each page was then stored in markdown format within a consolidated CSV file. We applied the same structured prompting and extraction workflow to these segments to ensure consistency and completeness of the final output. 

This revised workflow is designed to handle diverse document types, including API-accessed articles, standard and scanned PDFs, and historical handbooks. It enables reliable extraction of scientific content into structured, machine-readable formats. Finally, using this pipeline, we built an automated database of magnetic materials comprising 67,573 entries.

\subsection*{Feature Engineering}
A feature is the numerical representation of data used as input for the model. They are represented in the multi-dimensional space as vectors, $X = (x_1, x_2, \ldots, x_n) \in \mathbb{R}^n$. It encapsulates all central aspects of data that are relevant to the target variable. Feature engineering generates the most appropriate features based on the specific data, model, and task. This is generally one of the critical steps in developing a predictive model, as it dictates how well and effectively such a model can perform. The importance of feature engineering is more prominent in materials science, where the relationships among material composition, structure, and properties are complex. Proper feature engineering in this domain thus calls for a delicate balance between domain expertise and data-driven insight in order to effectively capture the underlying physics and chemistry regulating the behavior of materials.

In this study, we have focused more on the chemical composition and structural properties to make a feature vector. From the chemical formula, we built up an elemental proportion vector for each compound that might represent the compositional complexity of materials. This vector gives a proportional indication of how each element is distributed in a material's chemical composition. First, we counted the total number of unique elements found in all the compounds in the dataset. In this case, we found 84 unique elements, which cover a big part of the periodic table. Then we generated an 84-dimensional vector for each compound where each component corresponds to one speciﬁc element. The elemental proportion $p_{i}$ for element i is defined as:

    \begin{equation}
    p_{i} = \frac{n_{i}}{N}
    \label{eq:probability}
\end{equation}
where $n_{i}$ is the number of atoms of element i in the compound, and N is the total number of atoms in the compound.
Thus, our elemental proportion vector is represented as:
\begin{equation}
    p = (p_1, p_2, p_3,......, p_{84})
\end{equation}

where $p_{i}$ take zero value if the element is not present in the material chemical composition.
We have also generated other features related to material using atomic properties. For example, we constructed the average atomic number of each compound with the atomic number of every element in it. We did this computation by summing all products of elemental proportion vectors with atomic numbers of corresponding vectors. Mathematically, 

\begin{equation}
    \text{Average Atomic Number}  (\bar{Z}) = \sum_{i=1}^{84} p_{i} Z_{i}
\end{equation}
where p$_{i}$ is the proportion of element i in the compound and Z$_{i}$ is the atomic number of element i. 
Similarly, we generated several additional features like average atomic weight, average electronegativity, average magnetic moment, average group, and average period of all compounds in the dataset\cite{ong2013python}.

Additionally, we calculated the L2 stoichiometry norm and Entropy from the chemical composition of the compound. We used the following formulas to calculate these features\cite{ward2016general, troparevsky2015criteria}.

\begin{equation}
    \text{L2 Stoichiometry Norm } (L2) = \sqrt{\sum_{i=1}^{84} p_{i}^{2}}
\end{equation}

\begin{equation}
    \text{ Entropy } (\bar{S}) = -\sum_{i=1}^{84} p_{i} \log{p_{i}}
\end{equation}

The last two features from the chemical composition are the total proportion of high Curie temperature magnetic elements like Fe, Co, and Ni, and a total proportion of the rare earth elements in the compound.

For the structure-informed model, we created two additional features from the structural details. 
Since the crystal system is a nominal categorical variable consisting of seven distinct types, assigning arbitrary numeric values (such as integers from 1 to 7) could result in the model incorrectly interpreting these values as having a specific order or hierarchy, which is not the case. To address this, we employed the one-hot encoding technique.
One-hot encoding is a method that converts categorical variables into a set of binary variables (0s and 1s), ensuring that the model does not assume any implicit ordering between the categories. Each type of crystal system was represented as a separate binary feature. For example, a crystal system type like “Cubic” would be represented as [1, 0, 0, 0, 0, 0, 0], while another type such as “Tetragonal” would be represented as [0, 1, 0, 0, 0, 0, 0]. This transformation allowed us to represent the categorical data in a numerically stable and unbiased way for the machine learning model.

In the case of the space group feature, which consists of more than 100 distinct categories in our case, using one-hot encoding was not feasible due to the high dimensionality it would introduce. The rest of the feature space has fewer than 100 features, so adding more than 100 binary variables for space groups would significantly increase the complexity of the model.
To address this, we applied label encoding. Instead of treating the space group as a categorical variable, we assigned a numerical value to each space group based on the average of the target variable for that group. This approach reduced the space group feature to a single dimension while preserving relevant information, making it computationally efficient and manageable within our model.

\subsection*{Model Development: Classification Model}

To classify materials as nonmagnetic (NM), ferromagnetic (FM), or antiferromagnetic (AFM), we trained and evaluated two machine learning classifiers: a RF classifier and an XGBoost classifier.
Random Forest is an ensemble learning technique that constructs multiple decision trees and outputs the majority class as the final prediction. It is known for robustness against overfitting and its ability to model complex, high-dimensional feature spaces.
Our classification task required distinguishing among three classes, but our database contained only magnetic materials (FM and AFM). To incorporate nonmagnetic examples, we augmented our dataset with 11,389 stable nonmagnetic compounds from the Materials Project database~\cite{jain2013commentary}, yielding a combined dataset of 35,037 labeled materials.

Features were generated from the chemical composition of each material, as described in the Feature Engineering section. The target variable \texttt{Type} was encoded numerically: 0 = FM, 1 = AFM, 2 = NM.
A stratified 60/20/20 split was performed to divide the data into training, validation, and test sets, ensuring balanced representation of all three classes. For both models, classification pipelines were constructed and optimized using grid search with five fold cross-validation, with accuracy as the evaluation metric.The hyperparameters tuned included the number of estimators, maximum tree depth, minimum samples required to split a node, minimum samples per leaf, and the number of features considered at each split. For the XGBoost model, additional parameters such as learning rate, subsample ratio, and column sampling ratio were also optimized. 

The Random Forest classifier was implemented using Scikit-learn~\cite{pedregosa2011scikit} with the following best-performing hyperparameters: \texttt{n\_estimators=500}, \texttt{max\_depth=None}, \texttt{min\_samp.\_split=2}, \texttt{min\_samp.\_leaf=1}, \texttt{max\_features=`sqrt'}. We also used \texttt{class\_weight=`balanced'} to account for class imbalance.The XGBoost classifier was trained using similar procedures, with the following tuned hyperparameters:\\ \texttt{n\_estimators=800}, \texttt{max\_depth=10}, \texttt{learning\_rate=0.1}, \texttt{subsample=0.7}, \texttt{colsample\_bytree=0.6}.

Both models were evaluated on the validation and test sets using accuracy, confusion matrices, and detailed classification reports. The model achieved similar levels of accuracy and classification performance on the training, validation, and test sets, indicating strong generalization and stability across different data splits.

\subsection*{Model Development: Regression Models}

In this section, we trained and evaluated several machine learning models---Random Forest (RF), Extreme Gradient Boosting (XGBoost), and Ensemble Neural Network (ENN)---to estimate the Curie and N\'eel temperatures of ferromagnetic and antiferromagnetic compounds. Separate datasets were prepared for each magnetic phase using compounds from our NEMAD database. The ferromagnetic dataset initially contained 24,500 entries, and the antiferromagnetic dataset had 9,500 entries. To ensure each chemical composition appeared only once, we grouped entries by composition. When multiple transition temperatures were reported for the same composition, we computed their mean to obtain a unique transition temperature. This preprocessing step resulted in datasets of 15,577 and 7,893 unique compounds, respectively.

The ferromagnetic dataset exhibited a highly skewed temperature distribution, with approximately 47\% of entries having Curie temperatures below 300~K and only 22\% above 600~K. To address this imbalance, we adopted a stratified undersampling approach. We manually divided the Curie temperature range into fixed bins and identified the bin with the smallest sample count. For each bin, we randomly sampled a fixed number of compounds equal to this minority count, ensuring balanced representation across bins. The resulting balanced dataset comprised 12,461  ferromagnetic and 6,314 antiferromagnetic compounds.
We also prepared a structure-informed dataset containing 5,066 ferromagnetic compounds, which included additional structural features like crystal system and space group.
For all models, we used input features derived from the chemical formula, as described in the feature engineering section. In the case of the structure-informed dataset, these features were augmented with structural descriptors.

We used three models (RF, ENN, and XGBoost) on both the original and balanced datasets. For the original dataset, we directly trained all models using a stratified 60/20/20 split into training, validation, and test sets, preserving the distribution of Curie temperatures through bin based stratification. For the balanced dataset, we combined this stratified splitting with undersampling and trained ensemble models to improve generalization and capture predictive uncertainty. Specifically, we constructed multiple balanced sets by randomly undersampling the overrepresented low temperature bins while retaining all high temperature compounds in each set. Each ensemble member was trained on a different balanced set, allowing the models to always see the full set of high temperature data and varying portions of the low temperature data. This ensured that all data points from the original dataset were utilized across the ensemble, preventing overfitting to the dominant low-temperature region while still learning from the complete dataset.

The performance of the models was evaluated using R$^2$, mean absolute error (MAE), and root mean square error (RMSE).

Hyperparameters were optimized using grid search combined with five-fold StratifiedKFold cross-validation. Stratification was based on binned Curie temperature values to preserve distribution across folds. For each hyperparameter configuration, we trained an ensemble of five models on undersampled subsets to ensure balanced representation across temperature ranges. The best hyperparameter configuration was chosen based on the average validation performance, calculated across all cross-validation folds and the ensemble models trained in each fold.
\noindent
The best hyperparameters for Random Forest were: 
\texttt{n\_estimators=1000}, 
\texttt{max\_depth=None}, 
\texttt{min\_samples\_split=2}, 
\texttt{min\_samples\_leaf=1}, and 
\texttt{max\_features='sqrt'}.

\noindent
The optimal hyperparameters for XGBoost were: \texttt{n\_estimators=1200}, \texttt{max\_depth=12}, \texttt{learning\_rate=0.08}, \texttt{subsample=0.8}, and \texttt{colsample\_bytree=0.6}.

For the final ensemble, we trained 30 models per algorithm on differently sampled subsets of the full training data and averaged their predictions. The standard deviation across these ensemble outputs was used to generate confidence intervals, providing a measure of uncertainty. This method was applied consistently across the original, balanced, and structure-informed datasets.

In the case of ENN, we first split the dataset into 80\% training and 20\% test sets using stratified sampling based on binned Curie temperatures. To assess model stability and generalization, we performed five-fold StratifiedKFold cross-validation on the training set. In each fold, we trained an ensemble of five neural networks using stratified undersampling to balance temperature ranges, and averaged their predictions to evaluate fold-level performance. 

For final testing, we trained an ensemble of 30 fully connected neural networks, each with 8 hidden layers and ReLU activation, using the Adam optimizer with a learning rate of 0.001. Each model was trained on a different stratified-undersampled subset of the full training data for 500 epochs with a batch size of 64. Final predictions were obtained by averaging the outputs of all ensemble members, while the standard deviation across predictions quantified uncertainty.

\section*{\textbf{Data Availability}}
The full Northeast Materials Database (NEMAD), including all curated magnetic compounds and their associated properties, is publicly accessible at \url{https://www.nemad.org}. 

\section*{\textbf{Code Availability}}

All source code used for the machine learning analysis, together with the processed datasets employed to train and evaluate the models, is available at \url{https://github.com/sumanitani/NEMAD-MagneticML}  and has been archived on Zenodo (\url{https://doi.org/10.5281/zenodo.17042814}) \cite{Itani2025NEMAD}.

\section*{Acknowledgements}

This work was supported by the Office of Basic Energy Sciences, Division of Materials Sciences and Engineering, U.S. Department of Energy, under Award No. DE-SC0020221. The authors thank Houssam Sabri for insightful and fruitful discussions that contributed to this work.
\section*{Author Contributions Statement}

J.Z. conceived the project. S.I. and Y.Z. built the materials database using LLMs. S.I. developed classification and regression models for materials prediction. Y.Z. developed the NEMAD website. S.I., Y.Z. and J.Z. wrote the manuscript.

\section*{\textbf{Competing Interests Statement}} 
The authors declare no competing interests.

\clearpage

\clearpage

\end{document}